\documentclass[12pt]{article}

\usepackage{graphics}
\usepackage{amsmath}
\def\ni{\noindent}
\def\ph{{\phantom{...}}}
\def\={\phantom{..} = \phantom{..}}

\def\+{\phantom{..} + \phantom{..}}
\def\>{\phantom{..} > \phantom{..}}
\def\<{\phantom{..} < \phantom{..}}
\def\-{\phantom{..} - \phantom{..}}

\def\bq{\begin{quote}}
\def\eq{\end{quote}}
\def\be{\begin{equation}}
\def\ee{\end{equation}}
\def\bar{\begin{eqnarray}}
\def\ear{\end{eqnarray}}
\def\no{\nonumber}

\def\Sch{Schr{\"o}dinger}
\def\Schism{Schr{\"o}dingerism}
\def\Schist{Schr{\"o}dingerist}
\def\Schists{Schr{\"o}dingerists}

\def\Copism{Copenhagenism}
\def\Copist{Copenhagenist}
\def\Copists{Copenhagenists}

\def\vN{von Neumann}

\def\tr{\hbox{tr}}

\def\cH{{\cal H}}
\def\cO{{\cal O}}

\def\Eth{{\cal E}_{\hbox{th.}}}
\def\oort{\frac{1}{\sqrt{2}}}

\title{\bf Can the Infamous Boundary Be Found in Macromolecules?\\[0.5in]
 also: \vN\ {\em vs}. \Sch\ ensembles, and `Hund's Paradox' in quantum chemistry\\[3in]}

\author{W. David Wick\footnote{email: wdavid.wick@gmail.com}}

\begin{document}
\maketitle
\pagebreak

\section*{Abstract}

John Bell coined the phrase ``Infamous Boundary" for the point where classical physics splits off
from quantum physics. Many authors, including the present one, have advanced theories 
with the intention of defining and locating this ``shifty split"; 
most propose that it lies somewhere
on the scale of apparatus. But what if it resides at the level of macromolecules?
I show here that this question is intimately connected to the choice of thermal ensembles
and to the so-called `Hund's Paradox' in quantum chemistry. I propose an experimental
set-up that could in principle reveal the IB lurking in asymmetric macromolecules.

\section{Introduction\label{introsection}}

In January 2025, I read an obituary in the New York Times\footnote{19 January 2025, p. A 28.}
 of the Nobel-Prize-winning chemist,
Dr. Martin Karplus. The headline referred to Karplus as the man ``Who Made Computers
a Chemist's Tool", and contained this sentence: 

\begin{quote}

The models used classical Newtonian physics to predict how multitudes of atoms 
and molecules move during reactions, and they used quantum physics to describe 
how chemical bonds are broken and formed
during these reactions. This type of analysis proved particularly useful 
in understanding biological reactions involving enzymes, 
the proteins that govern chemical responses in living organisms.

\end{quote}

Why, I wondered, could classical physics serve for modeling some molecules, or parts thereof,
while ``quantum physics" was necessary for other molecules or parts? Where was the dividing line?

In this paper I present an argument suggesting (certainly not proving) that this line---perhaps
identical with the Infamous Boundary of John Bell---might  
be found in large organic molecules (macromolecules). The argument will derive
from various considerations, which will be presented in separate sections. 
An experimental program will also be described
in a later section, in order to evade the charge that 
I am merely presenting an ideological polemic. A Discussion section at the end contains some
historical anecdotes and some reflections on rival paradigms.

\section{A Clash of Ensembles\label{ensemblessection}}

A {\em von Neumann Thermal Ensemble} (vNTE) is defined as follows, \cite{vNbook}. Let $\cO$
denote an observable---which, for a \Copist, means a self-adjoint linear operator 
(matrix) on the Hilbert 
space of the system---and `$\cH$' the Hamiltonian---another self-adjoint operator.
Then:

\bar
\no \Eth\,\left[\,\cO\,\right] &\=& \tr\,\left[\,\exp\{\,- \cH/kT\,\}\,\cO\,\right]/Z;\\
\no Z &\=& \tr\,\left[\,\exp\{\,- \cH/kT\,\}\,\right].\\
&&
\ear

\ni Here `$\Eth$' on the left stands for ``thermal expectation", `$k$' is Boltzmann's
constant, and `$T$' is the temperature. Now assume that everything is in a finite box
and so $\cH$ has purely discrete spectrum:

\be
\cH\,\psi_n \= \lambda_n\,\psi_n, \ph \hbox{for}\ph n = 0,1,2,...
\ee

\ni Then:

\bar
\no \Eth\,\left[\,\cO\,\right] &\=& \sum_{n=0}^{\infty}\,<\psi_n\,|\cO|\,\psi_n>
\,\exp\{\,- \lambda_n/kT\,\}/Z;\\
\no Z &\=& \sum_{n=0}^{\infty}\,
\,\exp\{\,- \lambda_n/kT\,\}.\\
&&
\ear

This formula is exactly what we would write if we assumed (a) an ensemble member
is always in an eigenstate of $\cH$; (b)  each eigenstate has frequency in the ensemble
given by the Gibbs factor (the factor with the exponential); 
and (c) if it is in state $\psi_n$, then the ``expected
value" of $\cO$ is $<\psi_n\,|\cO|\,\psi_n>$. Part (c) here reflects the \Copist\ belief
that wavefunctions, or (for those who cannot abide them) 
Hilbert space vectors, are stand-ins for catalogs of probabilistic
expectations (which \Schists\ of course reject). 

Note that {\em there are no superpositions of energy states in a} vNTE.

A {\Schist\, thermal ensemble} (STE) is defined as follows. Let `$\cO_S$' be
an observable---which means for \Schists\ a {\em functional of the wavefunction}.
Then:

\bar
\no \Eth\,\left[\,\cO\,\right] &\=& \int_{[\sum |a_n|^2 = 1]}\,\prod\,dx_n\,dy_n\,
\exp\{\,- \cH_S(\psi)/kT\,\}\,\cO_S(\psi)/Z;\\
\no Z &\=& \int_{[\sum |a_n|^2 = 1]}\,\prod\,dx_n\,dy_n\,
\exp\{\,- \cH_S(\psi)/kT\,\}.\\
&&\label{ste}
\ear

\ni Here $\psi = \sum\,a_n\,\psi_n$, with $a_n = x_n + i\,y_n$. 
$\cH_S$ is another functional of $\psi$ which,
in linear wavefunction theory (meaning \Sch's theory of 1926) would be written:

\be
\cH_S(\psi) \= <\psi\,|\cH|\,\psi>,
\ee

\ni but is {\em not} the ``expected value of the energy" but simply ``the energy"
of state $\psi$. However, we do not insist that $\cH$ take this form.

Note that a STE is made up of {\em superpositions}; the probability of an ensemble member
being an {\em exact eigenstate} of $\cH$ is exactly zero.

\section{Symmetry-Breaking Molecules and `Hund's Paradox'\label{moleculessection}}

Chemists know well that molecules with the identical elemental makeup and chemical formula
may have very different activities. For example, many molecules of biological importance,
including proteins, DNA, and enzymes, possess left-handed forms which cannot be superimposed
on the right-handed versions (it is said they differ in ``chirality") 
and for which, say, the left-handed form is biologically active 
while the right-handed molecule is biologically inert. They are called ``enantiomers" of
the molecular recipe. There are also ``isoforms" resulting from symmetry-breaking events.

Let `$A$' and `$B$' represent enantiomers or isoforms that break a symmetry of the Hamiltonian.
Let $\psi_A$ and $\psi_B$ denote wavefunctions concentrated on these forms. Early on,
it was noticed (perhaps by Hund in 1927, \cite{Hund}) that neither can be the ground state,
i.e., the wavefunction with the lowest possible energy. 
One reason often cited is that these wavefunctions
will have some overlap in space and so are not orthogonal in the Hilbert space---so one
cannot be the ground state and the other an excited state. However, these states could be:

\bar
\no \psi_0 &\=& \oort\,\psi_A + \oort\,\psi_B;\\
\no && \\
\no \psi_1 &\=& \oort\,\psi_A - \oort\,\psi_B.\\
&&
\ear

We can turn around these formulas to give:

\bar
\no \psi_A &\=& \oort\,\psi_0 + \oort\,\psi_1;\\
\no && \\
\no \psi_B &\=& \oort\,\psi_0 - \oort\,\psi_1.\\
&&
\ear

Now to ``Hund's Paradox", which is paradoxical only for a \Copist: a {\em vNTE will contain no 
members equal to either $\psi_A$ or $\psi_B$} (because it contains no superpositions of
eigenstates). Yet we ``see" {\em either} $A$ or $B$ forms and never something in between.

While for an STE, noting that

\be
\cH(\psi_A) = \cH(\psi_B) > \cH(\psi_0),
\ee

\ni if the temperature is positive some ensemble members will contain superpositions
more or less weighted on one of them.  

\section{What is Meant by ``Seeing" an Isoform?\label{seeingsection}}

We do not ``see" atoms or molecules; we see device registrations (needles moving on a dial,
lights flashing, that sort of thing). In Bohr's philosophy, devices are ``classical"; hence,
they cannot enter into superpositions. In this author's 2017 proposal, an additional
contribution to the energy beyond the usual expression was assumed; call it $WFE$. It
took the form:

\be
WFE(\psi) = w\,N^2\,D_N(\psi),
\ee

\ni where `$w$' is a positive parameter (of unknown magnitude) and `$N$' is the number of degrees
of freedom of system+apparatus. The last factor is a dispersion, computed from the wavefunction,
of the center-of-mass (COM) of the combined system. The dynamics derived from the resulting
Hamiltonian is nonlinear but retains many good properties of the linear case, see \cite{WickI}.
At that time, I assumed that $WFE$ becomes significant only on the level of macroscopic devices,
i.e., with $N \approx 10^{20}$ or larger. Hence the parameter `$w$' could be taken
so small that at the atomic level the effects of incorporating 
$WFE$ would be infinitesimal, explaining
why linear equations work so well on that scale.

In \Schism, and perhaps \Copism\ too, the combined system must be described by a joint
wavefunction. I'll denote those by: $\Psi$. If we start with the microsystem in the superposition:

\be
\psi_0 \= \oort\,\psi_A + \oort\,\psi_B,
\ee

\ni then assuming the usual linear dynamics the combined system must evolve into:

\be
\Psi_0 \= \oort\,\Psi_A + \oort\,\Psi_B,
\ee

\ni where $\Psi_A$ is concentrated on configurations in which the molecule
has form $A$ and the device points to `$A$ was detected", and similarily for
$B$ replacing $A$. Hence, we have no outcome yet (this is the gist of the Measurement Problem).

In \Copism, some outside influence (for von Neumann it was ``the consciousness of the observer")
comes in and collapses the wavefunction to be either $\Psi_A$ or $\Psi_B$. 
In this author's 2017 theory, assuming that $\Psi_A$ and $\Psi_B$ have some relative 
displacement of the COM (say of some apparatus part), it may be that the energies 

\be
WFE(\Psi_0) \ph \hbox{and} \ph WFE(\Psi_1)
\ee

\ni are impossibly large (for terrestrial laboratories to achieve, anyway).
For the nonlinear dynamics there are then two possibilities: (a) nothing happens
(no registrations), or (b) the combined system is forced to make a choice and then
approach state $\Psi_A$ or $\Psi_B$ (with, e.g., a device pointer moving toward showing
``$A$ registered" or ``$B$ registered"). Some small-scale simulations indicated that (b)
is correct, \cite{WickIII}. The `decision': $A$ or $B$? may then be set by initial
conditions and may appear random (think ``chaos"), \cite{Wickchaos}. 

Thus, if we are only concerned with what we can perceive with our own senses on devices, 
both paradigms (\Copist\ and \Schist\ with nonlinear modification) can explain
why we see only $A$ {\em or} $B$. That is because each offers a putative solution of
the Measurement Problem.
As two philosophers of chemistry, Alexander Franklin and Vanessa A. Seifert,
noted, this means, as they announced in the title of an article, \cite{FandS}:
 ``The Problem of Molecular Structure Just is the Measurement Problem".

\section{Where Lies the Infamous Boundary?\label{wheresection}}

This author assumed in his 2017 that $WFE$ becomes significant only for macroscopic
apparatus. What if it is functional at the level of macromolecules?

Suppose

\be
\left\{\,WFE(\psi_A), WFE(\psi_B)\,\right\} <<
\left\{\,WFE(\psi_0), WFE(\psi_1)\,\right\}
\ee

\ni because $\psi_0$ and $\psi_1$ have `large' dispersions of their COM.
Suppose as a consequence

\be
<\psi_A\,|\cH|\,\psi_A> + WFE(\psi_A)\ph < \ph
<\psi_0\,|\cH|\,\psi_0> + WFE(\psi_0),
\ee

\ni even though

\be  
<\psi_A\,|\cH|\,\psi_A> \ph > \ph
<\psi_0\,|\cH|\,\psi_0>, 
\ee

\ni with the same relations holding with $B$ replacing $A$.

In this case, call it ``case I", sufficient cooling might put each ensemble member into
{\em either} $\psi_A$ or $\psi_B$; i.e., the ensemble would become
a statistical mixture of the two forms.

By contrast, suppose 

\be
<\psi_A\,|\cH|\,\psi_A> + WFE(\psi_A)\ph > \ph
<\psi_0\,|\cH|\,\psi_0> + WFE(\psi_0),
\ee

\ni and the same with $B$ replacing $A$. In this case, call it ``case II",
 the cooled ensemble might contain superpositions
of the two forms.

\section{Detecting the Superposition\label{detectingsection}}

Now suppose our molecule is an enzyme that can act on a substrate. Suppose also that
isoform $A$ can convert the substrate, but not isoform $B$. 
Then in case I of the previous section we might find the substrate either completely
converted or unaffected. While in case II, presumably one-half of the substrate
might be converted.

I have to leave inventing a practical plan for performing such an experiment to the chemists.
Certainly it will be difficult to both cool a molecule nearly to its ground state and
also watch whether it participates in a chemical reaction.

For enantiomers, another possibility may be to shine linearly polarized 
light on the system,
because the two forms rotate the plane of polarization in opposite directions. 
In case I, presumably
outgoing light would be in a linearly polarized state (or a mixture of two such),
while in case II we would observe a superposition of these states (which may generate
circularly-polarized light). 

\section{Discussion\label{discussionsection}}

We know that molecules of reasonably-large size (e.g., 2000 atoms) but still small with respect
to proteins or DNA double-helices can be put into superpositions (and even sent through
interferometers
which revealed fringes, \cite{NatPhys}). Thus the research agenda put forward in
the last section, which might involve laboratory work or a literature search,
can be recast as: find molecules which can or cannot be put into superpositions,
and then rank them by physical size and dispersion of the COM in the ground state.
Display the result of this search in a figure or table.

Given success in this endeavor, it would interesting next to ask whether
reaction centers of macromolecules lie on the ``quantum side" (meaning where
superpositions are possible), while other parts reside in the ``classical side".

The author's theory from 2017 unfortunately contains a free parameter 
I wrote as `$w$'.\footnote{It was the only Roman lower-case letter not
already assigned in the physics literature and not used by him in other sections.} 
If the experimental program described in section \ref{detectingsection} 
is actually carried out,
the data generated may allow for an estimate of the magnitude of `$w$'.  

Now to historical/ideological/paradigmatical matters. 
Although I should not presume to speak for a rival faction, I suspect that
\Copists\ would argue that the explanation for why case I of section \ref{wheresection}
yields a statistical mixture is that the molecule performed an observation on itself.
There are some precedents; for instance, about a decade ago I encountered articles about ``quantum
thermodynamics" in which, to solve the problem of superpositions, the authors proposed
that, in a large system, one-half of it could be regarded as observing the other half,
and so we can collapse the latter's wavefunction (although they
did not say which halves were the observers and the observed, or how often this collapse
occurs). In more distant times, in the celebrated debate over why the proton was not
decaying in experiments (some ``Grand Unified Theories" proposed that it would),
some authors argued that the proton was stable only because it is under continuous
observation by the neutrons (and so they could invoke the Quantum Zeno's Paradox, also known as
the ``A Watched Pot Never Boils Paradox"), \cite{TIB}\footnote{pp. 168-170 of the paperback
edition.} 

Besides self-observation, \Copists\ have offered many other resolutions for
Hund's Paradox: slow tunneling between the two enantiomers; decoherence, due to
a (rarely described, mysterious) ``environment", a small parity-violating energy; 
and many others, see e.g., \cite{FandS}
for references. For every anomaly in a paradigm, there are always many ``outs"
that avoid our having to abandon it.

Concerning that Clash of Ensembles (section \ref{ensemblessection}): \Sch\ himself vacillated.
Here is a curious forward to a book about thermodynamics
 \Sch\ published in 1952, \cite{Sbook}, but derived from lectures he gave in 
Dublin in 1944:

\begin{quote}

NOTE ON SECOND EDITION

The view that a physical process consists of continual jumplike transfers 
of energy parcels between microsystems cannot, when given serious thought, 
pass for anything but a sometimes convenient metaphor. 
To ascribe to every system always one of its sharp energy values 
is an indefensible attitude.  It was challenged in the beginning of Chapter II, 
yet it was adopted throughout this treatise as a convenient shortcut. 
The Appendix added to the Second Edition contains the general proof, 
that a consistent procedure, based on very simple assumptions, 
always gives the same results. The thermodynamical functions 
depend on the quantum-mechanical level scheme, not on the gratuitous allegation 
that these levels are the only allowed states. 

\end{quote}

What happened in those eight years, that caused this outburst? 
Was it \Sch's subconscious mind rebelling, 
or did fellow \Schists\ make remarks about intellectual consistency? 
But his Appendix can only be described as special pleading. 
For his ``general proof" he assumes at the outset that the entropy 
is given by the logarithm of the number of modes of a given energy, 
which is sensible only if you agree that systems actually reside in one such mode, 
which maligns the possibility of superpositions.  
(Demonstrating that his procedure is ``consistent"---with what?---doesn't persuade.)  
Thus \Sch\ assumes his desired conclusion. 
It's circular apologetics for, astonishingly, Copenhagenist doctrine. 

L. de Carlo and I found instances, \cite{deCarloWick},
 showing clearly that these superpositions matter; 
e.g., in the wavefunction version of a classical spin model of a magnet,
with the usual Hamiltonian,
no magnetization appears, while a Copenhagenist version 
recurs to the classical case and so magnetizes. But ditto for
a \Sch\ version that incorporates WFE. In \Sch's\ defense, we should
remark that we found the integrals in (\ref{ste}) very hard to perform,
and so we had to resort to approximations derived from probability theory. 

Perhaps we should just reflect on \Sch\ as a man of his times, 
unable to pursue against overwhelming opposition the implications 
of his revolutionary discovery of 1926. Or perhaps he had agreed to teach 
the stat mech course at the Dublin Institute, 
and, for the sake of the student's careers, felt he had to toe the party line  
as it existed at that time (and still does today)?

\section*{Acknowledgement} 
The author thanks L. de Carlo for valuable discussions.


\begin{thebibliography}{9}

\bibitem{vNbook}
von Neumann, J. {\em The Mathematical Foundations of Quantum Mechanics}.
 English translation of the original German text (1932), 
published by Princeton University Press (1955). 

\bibitem{Hund}
Hund, F. ``On the Explanation of Molecular Spectra, II", Zeitschrift fur Physik, 42, 43-120 
(1927).

\bibitem{WickI}
Wick, W. D. ``On Non-Linear Quantum Mechanics and the Measurement Problem I: Blocking Cats", 
ArXiv 1710.03278 (October 2017).

\bibitem{WickIII}
Wick, W. D. ``On Non-Linear Quantum Mechanics and the Measurement Problem
III: Poincar{\' e} Probability and ... Chaos?" ArXiv 1803.1126v1 published March 2018.

\bibitem{Wickchaos}
Wick, W. D. ``Chaos in a Nonlinear Wavefunction Model: An Alternative to Born's Probability
Hypothesis". ArXiv 2502.02698. (4 Feb. 2025).


\bibitem{FandS}
Franklin, A and Seifert, V. A.
 ``The Problem of Molecular Structure Just is the Measurement Problem".
Brit J Phil Science, 75(1) 2024, also available over the Internet.

\bibitem{NatPhys}
Fein, Y. Y. {\em et al.} ``Quantum superposition of molecules beyond 25 kDa".
Nat Phys. 15, 1242-1245 (2019).

\bibitem{TIB}
Wick, W. D. {\em The Infamous Boundary: Seven Decades of Heresy in Quantum Physics.} 
With a mathematical appendix by William Faris. Birkh{\" a}user, 1995 (hard-bound,
with a different subtitle) and 
Copernicus, 1996 (paperback, with an index).

\bibitem{Sbook}
\Sch, E. {\em Statistical Thermodynamics}. 2nd edition (1952); 
reprinted, except for a Note to the reader and a new Appendix, 
from a “hectograph” of a lecture he gave in 1944 in Dublin.


\bibitem{deCarloWick}
De Carlo, L. and Wick, W. D. ``On \Schist\ Quantum Thermodynamics". ArXiv 2208.07688 (2022, 
revised in 2024).
Journal publication: ``On Magnetic Models in Wavefunction Ensembles". Entropy 25(4) 564 (2023).






\end{thebibliography}
\end{document}